\documentclass[
  twocolumn,
  prb,
  showpacs,
  amsmath,
  amssymb,
  superscriptaddress,
  floatfix
]{revtex4}

\usepackage{bm}
\usepackage{bbold}
\usepackage{graphicx}

\begin{document}

\title{Full counting statistics in disordered graphene at Dirac point:\\
From ballistics to diffusion}
\author{A.~Schuessler}
\affiliation{
 Institut f\"ur Nanotechnologie, Karlsruher Institut f\"ur Technologie,
 76131 Karlsruhe, Germany.
}

\author{P.~M.~Ostrovsky}
\affiliation{
 Institut f\"ur Nanotechnologie,
 Karlsruher Institut f\"ur Technologie,
 76131 Karlsruhe, Germany.
}
\affiliation{
 L.~D.~Landau Institute for Theoretical Physics RAS,
 119334 Moscow, Russia.
}

\author{I.~V.~Gornyi}
\affiliation{
 Institut f\"ur Nanotechnologie,
 Karlsruher Institut f\"ur Technologie,
 76131 Karlsruhe, Germany.
}
\affiliation{
 A.~F.~Ioffe Physico-Technical Institute,
 194021 St.~Petersburg, Russia.
}
\affiliation{
 DFG Center for Functional Nanostructures,
 Karlsruher Institut f\"ur Technologie,
 76131 Karlsruhe, Germany.
}

\author{A.~D.~Mirlin}
\affiliation{
 Institut f\"ur Nanotechnologie,
 Karlsruher Institut f\"ur Technologie,
 76131 Karlsruhe, Germany.
}
\affiliation{
 DFG Center for Functional Nanostructures,
 Karlsruher Institut f\"ur Technologie,
 76131 Karlsruhe, Germany.
}
\affiliation{
 Inst. f\"ur Theorie der Kondensierten Materie,
 Karlsruher Institut f\"ur Technologie,
 76131 Karlsruhe, Germany.
}
\affiliation{
 Petersburg Nuclear Physics Institute,
 188300 St.~Petersburg, Russia.
}

\begin{abstract}
The full counting statistics of the charge transport through an undoped graphene
sheet in the presence of smooth disorder is studied. At the Dirac point both in
clean and diffusive limits, transport properties of a graphene sample are
described by the universal Dorokhov distribution of transmission probabilities.
In the crossover regime, deviations from universality occur which can be studied
analytically both on ballistic and diffusive sides. In the ballistic regime, we
use a diagrammatic technique with matrix Green functions. For a diffusive
system, the sigma model is applied. Our results are in good agreement with
recent numerical simulations of electron transport in disordered graphene.
\end{abstract}

\pacs{73.63.-b, 73.22.-f}

\maketitle

\section{Introduction}

Electron transport in graphene remains a field of intense experimental and
theoretical activity \cite{Geim07, RMP07}. The hallmark of graphene is the
massless Dirac character of low-energy electron excitations. This gives rise to
remarkable physical properties of this system distinguishing it from
conventional two-dimensional metals. The most remarkable effects arise when the
chemical potential is located in a close vicinity of the neutrality (Dirac)
point. In particular, a short-and-wide sample  (with width $W$ much exceeding
the length $L$) of clean graphene exhibits at the Dirac point pseudo-diffusive
charge transport, \cite{Katsnelson} with ``conductivity'' $4e^2/\pi h$  and with
counting statistics (characterizing fluctuations of current) equivalent to that
of a diffusive wire. \cite{Dorokhov83, Tworzydlo06, Beenakker08rev, Ludwig07} In
particular, the Fano factor (the shot noise power divided by the current) takes
the universal value $F = 1/3$ that coincides with the well-known result for a
diffusive metallic wire. \cite{BeenakkerRMP} This is at odds with usual clean
metallic systems, where the conductance (rather than conductivity) is
independent of $L$ and the shot noise is absent ($F=0$). The reason behind these
remarkable peculiarities of transport in clean graphene at the Dirac point is
linearly vanishing density of states. This implies that the current is mediated
by evanescent rather than propagating modes. The above theoretical predictions
have been confirmed in measurements of conductance and noise in ballistic
graphene flakes. \cite{Morpurgo, Danneau, Miao07} Recent advances in preparation
and transport studies of suspended graphene samples also indicate that the
system may be in the ballistic regime. \cite{Du08, KimFQHE}

Effects of impurities on transport properties of graph\-ene are highly unusual
as well. In contrast to conventional metals, ballistic graphene near the Dirac
point conducts better when potential impurities are added.\cite{Titov07,
Bardarson07, Schuessler09} Quantum interference in disordered graphene is also
highly peculiar due to the Dirac nature of the carriers. In particular, in the
absence of intervalley scattering, the minimal conductivity\cite{Novoselov05}
$\sim e^2/h$ is ``topologically protected'' from quantum localization.
\cite{OurPapers} The exact value of the conductivity of such a system at the
Dirac point depends on the type of intravalley scattering (random scalar or
vector potential, or random mass, or their combination). For the case of random
potential only (which is experimentally realized by charged scatterers) the
conductivity in fact increases logarithmically with the length $L$, in view of
antilocalization. \cite{Bardarson07, Nomura07, San-Jose07, Lewenkopf08,
Tworzydlo08}

In our previous work\cite{Schuessler09}, we have studied the evolution of
conductance of a short-and-wide graphene sample from the ballistic to the
diffusive regime. We have also shown that the leading disorder-induced
correction to the noise and full counting statistics in the ballistic regime is
completely governed by the renormalization of the conductance. This implies, in
particular, that the Fano factor $1/3$ remains unaffected to this order. Indeed,
the experiments\cite{Danneau, Marcus08} give Fano factor values in the vicinity
of 1/3 at the Dirac point for different system lengths $L$. One could thus ask
whether deviations from this value should be expected at all.

In this work we present a detailed analysis of the shot noise and the full
counting statistics in samples with long-range (no valley mixing) disorder. We
show that to second order in the disorder strength a correction to the universal
counting statistics of the ballistic graphene does arise. We calculate this
correction and demonstrate that it suppresses the Fano factor below the value
1/3. For the case of random scalar potential, we also analyze the opposite limit
of large $L$ when the system is deep in the diffusive regime. Generalizing the
analysis of weak-localization effects on the counting statistics by
Nazarov\cite{Nazarov95}, we find that the Fano factor returns to the value of
1/3 from below with increasing $L$. The approach to 1/3 is however
logarithmically slow, These results compare well with recent numerical works
\cite{San-Jose07, Lewenkopf08} and particularly with the most detailed study by
Tworzydlo {\it et al.}\cite{Tworzydlo08}

The structure of the paper is as follows. In the Sec. \ref{sec:mgfm}, we
describe the general matrix Green function formalism and its application to the
problem of full counting statistics. The model for graphene setup and disorder
is introduced in Sec.\ \ref{sec:model}. We proceed with applying matrix Green
function method to the calculation of the distribution of transmission
probabilities of the clean graphene sample in Sec.\ \ref{sec:clean}. In Sec.\
\ref{sec:ball} we evaluate perturbative disorder corrections to the full
counting statistics in ballistic regime. Diffusive transport through disordered
graphene is considered in Sec.\ \ref{sec:dl} within the sigma-model approach.
The paper is concluded by Sec.\ \ref{sec:sum} summarizing the main results.
Technical details of the calculation are presented in three appendices.

\section{Matrix Green function formalism}
\label{sec:mgfm}

We begin with the general presentation of the matrix Green function approach to
the full counting statistics of a quasi-one-dimensional system. This formalism
was developed by Nazarov in Ref.\ \onlinecite{Nazarov94}.

Consider a quasi-one-dimensional sample attached to two perfect metallic leads.
Transport characteristics of the system are encoded in the matrix of
transmission amplitudes $t_{mn}$, where the indices enumerate conducting
channels (quantized transverse modes) in the leads. Eigenvalues of the matrix
$\hat t^\dagger \hat t$ determine transmission probabilities of the system (we
use the ``hat'' notation for matrices in the space of channels). Our main goal
is to calculate the distribution of these transmission probabilities. The full
counting statistics of the charge transport is given by the moments of this
distribution or, equivalently, by the distribution itself. The first two moments
of the transferred charge provide the conductance (by Landauer formula) and the
Fano factor
\begin{equation}
 G
  = \frac{e^2}{h} \mathop{\mathrm{Tr}} \hat t^\dagger \hat t, \\
 \qquad
 F
  = 1 - \frac{\mathop{\mathrm{Tr}} (\hat t^\dagger \hat t)^2}
      {\mathop{\mathrm{Tr}} \hat t^\dagger \hat t}.
 \label{GF}
\end{equation}

The starting point of our consideration is the relation between the matrix of
transmission probabilities and the Green function of the system,
\begin{equation}
 t_{mn}
  = i \sqrt{v_m v_n} G^A_{mn}(x, x'),
 \label{tG}
\end{equation}
Here $v_{m,n}$ are velocities in the $m$th and $n$th channel. The Green function
is taken in the mixed representation with real-space coordinates in the
longitudinal direction and channel indices in transverse direction. The
positions $x$ and $x'$ should be taken in the left and right lead respectively
in order to obtain the transmission matrix of the full system. The conjugate
matrix $\hat t^\dagger$ is related to the retarded Green function by a similar
identity.

The Green functions are defined in the standard way as
\begin{equation}
\label{eq:emHGeqd}
 (\epsilon - \hat H \pm i0) \hat G^{R,A}(x, x')
  = \delta(x-x') \hat{\mathbb{1}},
\end{equation}
with energy $\epsilon$ and Hamiltonian $\hat H$, the latter being an operator
acting both on $x$ and in the channel space.

With the help of Eq.\ (\ref{tG}) we can express all the moments of transmission
distribution in terms of the Green functions,
\begin{equation}
 \mathop{\mathrm{Tr}} \big( \hat t^\dagger \hat t \big)^n
  = \mathop{\mathrm{Tr}} \big[
      \hat v \hat G^A(x, x') \hat v \hat G^R(x', x)
    \big]^n.
 \label{L-C}
\end{equation}
where $\hat v$ is the velocity operator and $x$ and $x'$ lie in the left and
right lead respectively. For the first moment, $n = 1$, the above identity
establishes an equivalence of the Landauer and Kubo representations for
conductance.

The complete statistics of the transmission eigenvalues can be represented by
the generating function
\begin{equation}
 \mathcal{F}(z)
  = \sum_{n = 1}^\infty z^{n-1} \mathop{\mathrm{Tr}} \big(
      \hat t^\dagger \hat t
    \big)^n
  = \mathop{\mathrm{Tr}} \big[
      \hat t^{-1} \hat {t^\dagger}^{-1} - z
    \big]^{-1}.
 \label{Ffromtdt}
\end{equation}
Once this function is known, all the moments of transmission distribution are
easy to obtain by expanding the generating function in series at $z = 0$. An
efficient method yielding the whole generating function was proposed in Ref.\
\onlinecite{Nazarov94}. It amounts to calculating the matrix Green function
defined by the following equation:
\begin{multline}
 \begin{pmatrix}
   \epsilon - \hat H + i0 & -\sqrt{z} \hat v \delta(x - x_L) \\
   -\sqrt{z} \hat v \delta(x - x_R) & \epsilon - \hat H - i0
 \end{pmatrix} \check G(x, x') \\
  = \delta(x - x') \check{\mathbb{1}}.
 \label{matrixG}
\end{multline}
The parameter $z$ here corresponds to the source field mixing retarded and
advanced components of the matrix Green function. The positions $x_L$ and
$x_R$, where the source field is applied, lie within the left and right lead
respectively. We will refer to this specific matrix structure as the RA
(retarded -- advanced) space and denote such a matrices with the ``check''
notation.

The main advantage of the matrix Green function defined by Eq.\
(\ref{matrixG}) is the following concise expression for the generating
function of transmission probabilities:
\begin{multline}
 \mathcal{F}(z)
  = \frac{1}{\sqrt{z}} \mathop{\mathrm{Tr}} \left[
      \begin{pmatrix}
        0 & 0 \\
        \hat v & 0
      \end{pmatrix} \check G(x_R, x_R)
    \right]\\
  = \frac{1}{\sqrt{z}} \mathop{\mathrm{Tr}} \left[
      \begin{pmatrix}
        0 & \hat v \\
        0 & 0
      \end{pmatrix} \check G(x_L, x_L)
    \right].
 \label{F}
\end{multline}
The validity of this equation can be directly checked by expanding the Green
function in powers of $z$ with the help of perturbation theory and comparing
this expansion termwise with Eq.\ (\ref{Ffromtdt}). The equivalence of these
two expansions is provided by the identity (\ref{L-C}).

Another and, probably, most intuitive representation of the full counting
statistics is given by the distribution function of transmission probabilities
$P(T)$. This function takes its simplest form when expressed in terms of the
parameter $\lambda$ related to the transmission probability by $T = 1/\cosh^2
\lambda$. In terms of $\lambda$ the probability density is defined by the
identity $P(T) dT = P(\lambda) d\lambda$. The definition of the generating
function, Eq.\ (\ref{Ffromtdt}), implies a trace involving all transmission
probabilities. With the distribution function of these probabilities we can
express $\mathcal{F}(z)$ by the integral
\begin{equation}
 \mathcal{F}(z)
  = \int_0^\infty \frac{P(\lambda) d\lambda}{\cosh^2 \lambda - z}.
 \label{FfromP}
\end{equation}
The function $\mathcal{F}(z)$ has a branch cut discontinuity in the complex $z$
plane running from $1$ to $+\infty$. The jump of the function across the branch
cut determines the distribution function [see Ref.\ \onlinecite{Schuessler09}
for derivation]:
\begin{equation}
 P(\lambda)
  = \frac{\sinh 2\lambda}{2\pi i} \big[
      \mathcal{F}(\cosh\lambda + i0) - \mathcal{F}(\cosh\lambda - i0)
    \big].
 \label{PfromF}
\end{equation}
In other words, Eq.\ (\ref{FfromP}) solves the Riemann-Hilbert problem defined
by Eq.\ (\ref{PfromF}).

The generating function $\mathcal{F}(z)$ can be related to the ``free energy''
of the system in the ``external'' source field. The free energy is defined in
terms of the functional determinant
\begin{equation}
 \Omega
  = \mathop{\mathrm{Tr}} \ln \check G,
 \qquad
 \mathcal{F}
  = \frac{\partial \Omega}{\partial z}.
 \label{Omega}
\end{equation}
The free energy can be calculated by standard diagrammatic methods and hence
provides a very convenient representation of the full counting statistics. It
is also convenient to parametrize the argument of the free energy by the angle
$\phi$ according to $z = \sin^2(\phi/2)$.

Thus we have three equivalent representations of the full counting statistics by
the functions $\mathcal{F}(z)$, $P(\lambda)$, and $\Omega(\phi)$. In this paper
we calculate the transport characteristics of a disordered graphene sample in
terms of its free energy $\Omega(\phi)$. The two other functions can be found
with the help of identities
\begin{gather}
 \mathcal{F}(z)
  = \frac{2}{\sin \phi}\, \left.
      \frac{\partial \Omega}{\partial \phi}
    \right|_{\phi = 2 \arcsin\sqrt{z}}, \label{FfromO} \\
 P(\lambda)
  = \frac{2}{\pi} \mathop{\mathrm{Re}} \left.
      \frac{\partial \Omega}{\partial \phi}
    \right|_{\phi = \pi + 2 i \lambda}. \label{PfromO}
\end{gather}
The first of these relations directly follows from Eq.\ (\ref{Omega}) while the
second one is the result of the substitution of Eq.\ (\ref{FfromO}) into Eq.\
(\ref{PfromF}).

The two most experimentally relevant quantities contained in the full counting
statistics, namely, conductance and Fano factor, Eq.\ (\ref{GF}), can be
expressed in terms of any of the functions introduced above. Then the following
expressions for the conductance and the Fano factor hold:
\begin{gather}
 G
  = \frac{2 e^2}{h} \left.
      \frac{\partial^2 \Omega}{\partial \phi^2}
    \right|_{\phi = 0}, \label{Gdef}\\
 F
  = \frac{1}{3} - \frac{2}{3} \left.
      \frac{\partial^4 \Omega / \partial \phi^4}
           {\partial^2 \Omega / \partial \phi^2}
    \right|_{\phi = 0}. \label{Fdef}
\end{gather}

We will apply the matrix Green function formalism outlined in this section to
the problem of full counting statistics of a disordered graphene sample.
Our strategy is as follows. First, we calculate the matrix Green function
of a clean rectangular graphene sample and obtain the full counting statistics
with the help of Eq.\ (\ref{F}). Then we introduce disorder in the model
perturbatively. Evaluation of the free energy by diagrammatic methods yields
disorder corrections to the full counting statistics of a clean sample.

\section{Model}
\label{sec:model}

We will adopt the single-valley model of graphene. More specifically, we will
consider scattering of electrons only within a single valley and neglect
intervalley scattering events. Indeed, a number of experimental results show
that in many graphene samples the dominant disorder scatters electrons within
the same valley. First, this disorder model is supported by the odd-integer
quantization \cite{Novoselov05, Zhang05, Geim07} of the Hall conductivity,
$\sigma_{xy} = (2n+1) 2e^2/h$, representing a direct evidence\cite{OurQHE} in
favor of smooth disorder which does not mix the valleys. The analysis of weak
localization also corroborates the dominance of intra-valley
scattering\cite{Savchenko}. Furthermore, the observation of the linear density
dependence \cite{Geim07} of graphene conductivity away from the Dirac point can
be explained if one assumes that the relevant disorder is due to charged
impurities and/or ripples. \cite{Nomura07, Ando06, Nomura06, Khveshchenko,
OurPRB} Due to the long-range character of these types of disorder, the
intervalley scattering amplitudes are strongly suppressed and will be neglected
in our treatment. Finally, apparent absence of localization at the Dirac point
down to very low temperatures \cite{Novoselov05, Zhang05,Kim} points to some
special symmetry of disorder. One realistic candidate model is the long-range
randomness which does not scatter between valleys\cite{OurPapers,
footnote-chiral}.

The single-valley massless Dirac Hamiltonian of electrons in graphene has the
form (see, e.g., Ref.\ \onlinecite{RMP07})
\begin{equation}
 H
  = v_0 \bm{\sigma} \mathbf{p} + V(x,y),
 \qquad
 V(x,y)
  = \sigma_\mu V_\mu(x,y).
 \label{ham}
\end{equation}
Here $\sigma_\mu$ (with $\mu = 0,x,y,z$) are Pauli matrices acting on the
electron pseudospin degree of freedom corresponding to the sublattice
structure of the honeycomb lattice, $\bm{\sigma} \equiv \{\sigma_x, \sigma_y\}$,
and the Fermi velocity is $v_0 \approx 10^8$ cm/s. The random part $V(x,y)$ is
in general a $2 \times 2$ matrix in the sublattice space. Below we set $\hbar =
1$ and $v_0 = 1$ for convenience.

\begin{figure}
 \centerline{\includegraphics[width=0.9\columnwidth]{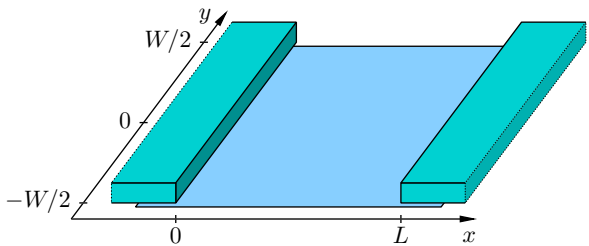}}
 \caption{(Color online) Schematic setup for two-terminal transport
measurements. Graphene sample of dimensions $L \times W$ is placed between two
parallel contacts. We assume $W\gg L$ throughout the paper.}
 \label{fig:sample}
\end{figure}

We will calculate transport properties of a rectangular graphene sample with
the dimensions $L \times W$. The contacts are attached to the two sides of the
width $W$ separated by the distance $L$. We fix the $x$ axis in the direction of
current, Fig.\ \ref{fig:sample}, with the contacts placed at $x=0$ and $x=L$.
We assume $W \gg L$, which allows us to neglect the boundary effects related to
the edges of the sample that are parallel to the $x$ axis (at $y=\pm W/2$).

Following Ref.\ \onlinecite{Tworzydlo06}, metallic contacts are modelled as
highly doped graphene regions described by the same Hamiltonian (\ref{ham}). In
other words, we assume that the chemical potential $E_F$ in the contacts is
shifted far from the Dirac point. In particular, $E_F \gg \epsilon$, where
$\epsilon$ is the chemical potential inside the graphene sample counted from the
Dirac point. (All our results are independent of the sign of energy, thus we
assume $\epsilon > 0$ throughout the paper.) We also assume zero temperature,
that is justified provided $T L \ll 1$.

With the boundary conditions specified above, we are able to calculate
explicitly the matrix Green function (\ref{matrixG}) for a clean graphene
sample [$V(x,y) = 0$] at zero energy. This calculation is outlined in Appendix
\ref{AppG} [see Eq.\ (\ref{G0})]. Using this Green function, we will study
disorder effects in the framework of the diagrammatic technique for the
free energy.

\section{Electron transport in clean graphene}
\label{sec:clean}

In this section we apply the matrix Green function formalism developed in Sec.\
\ref{sec:mgfm} to the case of clean graphene. These results will play the role
of the zeroth approximation for our perturbation theory.

The matrix Green function is derived in Appendix \ref{AppG}. The generating
function for the full counting statistics is given by Eq.\ (\ref{F}). With the
Green function (\ref{G0}), we obtain
\begin{equation}
 \mathcal{F}_0\big( \sin^2\tfrac{\phi}{2} \big)
  = \frac{W}{\sin\frac{\phi}{2}}  \mathop{\mathrm{Tr}} \left[
      \begin{pmatrix}
        0 & \sigma_x \\
        0 & 0
      \end{pmatrix} \check G(0, 0; 0)
    \right]
  = \frac{W}{\pi L} \frac{\phi}{\sin \phi}.
\end{equation}
The corresponding dependence of the free energy on the source field $\phi$
follows from integration of Eq.\ (\ref{FfromO}). This yields a simple
quadratic function
\begin{equation}
 \Omega_0(\phi)
  = \frac{W \phi^2}{4 \pi L}.
 \label{Omega0}
\end{equation}
This remarkably simple result reveals the convenience of the source field
parametrization $z = \sin^2(\phi/2)$. The clean sample responds linearly to the
``external'' field $\phi$. The distribution of transmission probabilities given
by Eq.\ (\ref{PfromO}) is just a constant, $P_0(\lambda) = W/\pi L$, in terms of
$\lambda$. This means the distribution acquires the Dorokhov form
\cite{Dorokhov83} characteristic for disordered metallic wires
\begin{equation}
 P_0(T)
  = \frac{W}{2 \pi L} \frac{1}{T \sqrt{1 - T}}.
\end{equation}
Hence electron transport in clean graphene at the Dirac point is often called
pseudodiffusive.

\begin{figure}
\centerline{\includegraphics[height=0.3\columnwidth]{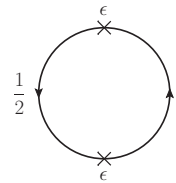}}
\caption{Lowest energy correction to the free energy of the system.}
\label{fig:eloop}
\end{figure}

Let us now calculate an energy correction to the pseudodiffusive transport
regime. In the vicinity of the Dirac point (we assume $\epsilon L \ll 1$), we
can account for finite energy $\epsilon$ by means of perturbation theory. The
linear term is absent due to particle-hole symmetry of the Dirac point. The
lowest non-vanishing correction appears in the $\epsilon^2$ order and is given
by the single diagram Fig.\ \ref{fig:eloop},
\begin{equation}
 \Omega_\epsilon
  = \frac{W\epsilon^2}{2}
    \int\limits_0^L dx\, dx' \int\limits_{-\infty}^\infty dy
    \mathop{\mathrm{Tr}} G(x, x'; y) G(x', x; -y).
 \label{Oe}
\end{equation}
This integral of the product of two Green functions is calculated in Appendix
\ref{AppE}. The result takes the form
\begin{equation}
 \Omega_\epsilon
  = \frac{W}{\pi L} \frac{(\epsilon L)^2}{\sin\frac{\phi}{2}}
    \frac{\partial}{\partial \phi} \left\{\!
      \cos\frac{\phi}{2} \left[
        \psi \left( \frac{\pi + \phi}{2\pi} \right)
        +\psi \left( \frac{\pi - \phi}{2\pi} \right)
      \right]
    \!\right\},
 \label{Oe_res}
\end{equation}
where $\psi$ is the digamma function.

As explained above, the free energy $\Omega_\epsilon$ contains information
about the full counting statistics, i.e. all moments of the transfered charge.
In particular, from Eqs.\ (\ref{Gdef}) and (\ref{Fdef}) we obtain the following
results for the conductance and Fano factor:
\begin{align}
 G
  &=\frac{4 e^2}{\pi h}\; \frac{W}{L} \left[
     1 + c_1 (\epsilon L)^2
   \right],
 \quad
 F
  =\frac{1}{3} \left[
     1 + c_2 (\epsilon L)^2
   \right],
 \label{GFsmallE}\\
 c_1
  &=\frac{35 \zeta(3)}{3 \pi^2} - \frac{124 \zeta(5)}{\pi^4}
  \approx 0.101,
 \label{c1}\\
 c_2
  &=-\frac{28 \zeta(3)}{15 \pi^2} - \frac{434 \zeta(5)}{\pi^4}
    +\frac{4572 \zeta(7)}{\pi^6}
  \approx -0.052.
 \label{c2}
\end{align}
These expressions coincide with the results of Ref.\ \onlinecite{Schuessler09}
obtained within an alternative (transfer-matrix) approach.

\section{Disordered graphene: ballistic limit}
\label{sec:ball}

Let us now include the random part $V(x, y)$ of the Hamiltonian (\ref{ham})
into consideration. There are in total four different types of disorder within
the single valley Dirac model: $V_0$ is the random potential (charged impurities
in the substrate), $V_x$ and $V_y$ correspond to the random vector potential
(e.g. long-range corrugations), and $V_z$ is the random mass. We will assume the
standard Gaussian type of disorder characterized by the correlation function
\begin{equation}
 \langle V_\mu(\mathbf{r}) V_\nu(\mathbf{r}') \rangle
  = 2 \pi \delta_{\mu\nu} w_\mu(|\mathbf{r} - \mathbf{r}'|).
 \label{dcorr}
\end{equation}
The functions $w_\mu(r)$ depend only on the relative distance $r$ and are
strongly peaked near $r = 0$. Thus we deal with isotropic and nearly white-noise
disorder. However, in order to accurately treat ultraviolet divergencies arising
in our calculation, we keep a small but finite disorder correlation length. The
results will be expressed in terms of four integral constants characterizing the
disorder strength
\begin{equation}
 \alpha_\mu
  = \int d\mathbf{r}\, w_\mu(|\mathbf{r}|).
 \label{amu}
\end{equation}

\begin{figure}
 \centerline{\includegraphics[height=0.27\columnwidth]{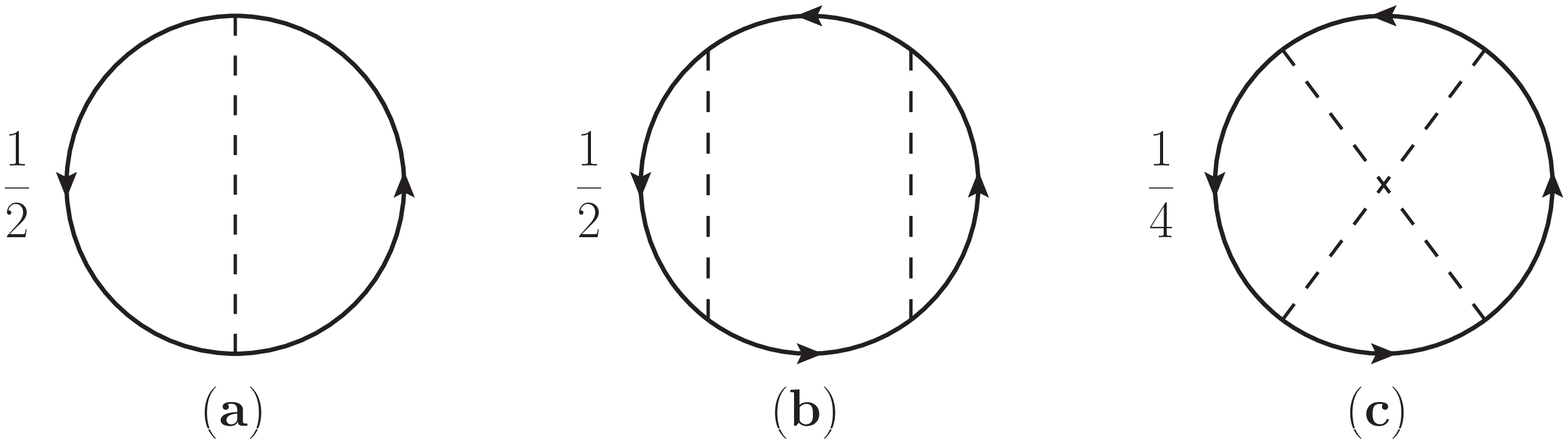}}
 \caption{
   Loop diagrams for the disorder corrections to the ground state energy
   $\Omega$, (a) first-order, (b) and (c) second-order.}
 \label{Fig:diagrams}
\end{figure}

Within the specified Gaussian disorder model, perturbative corrections to the
free energy are given by the loop diagrams. The first and second order
corrections are shown in Fig.\ \ref{Fig:diagrams}. Dashed lines in these
diagrams denote disorder correlation functions (\ref{dcorr}).

\subsection{First-order correction}

The first-order correction to the free energy $\Omega(\phi)$ is given by the
loop diagram containing two Green functions and one impurity line, Fig.\
\ref{Fig:diagrams}a. The Green function at coincident points diverges. That is
why we keep a finite correlation length calculating the first-order diagram.
Assuming the separation between two vertices of the diagram is given by the
vector $\bm{\delta}$, we obtain the expression
\begin{gather}
 \Omega_a
  = \int d\bm{\delta} \sum_\mu w_\mu(|\bm{\delta}|) \Omega_a^{(\mu)},
  \label{Omegaa} \\
 \delta \Omega_a^{(\mu)}
  = \pi \int d\mathbf{r} \mathop{\mathrm{Tr}} \Big[
      \sigma_\mu \check G(\mathbf{r}, \mathbf{r} + \bm{\delta})
      \sigma_\mu \check G(\mathbf{r} + \bm{\delta}, \mathbf{r})
    \Big].
\end{gather}
Now we substitute the Green function from Eq.\ (\ref{G0}) and expand the
correction to the free energy in powers of $\bm{\delta}$. For
the four possible disorder types, this yields
\begin{align}
 \Omega_a^{(0),(z)}
  &= \frac{\pi W}{2L^2} \int\limits_0^L dx \left[
      \frac{-1}{2 \sin^2 \frac{\pi x}{L}} \pm \left(
        \frac{\delta_x^2 - \delta_y^2}{6 \bm{\delta}^2}
        +\frac{\delta_y^2 \phi^2}{\pi^2 \bm{\delta}^2}
      \right)
    \right],\\
 \Omega_a^{(x),(y)}
  &= \frac{\pi W}{L^2} \int\limits_0^L dx \left[
      \frac{-1}{4 \sin^2 \frac{\pi x}{L}} \pm \left(
        \frac{1}{12}
        +\frac{\delta_x^2 - \delta_y^2}{\pi^2 \bm{\delta}^4}
      \right)
    \right].
\end{align}
In these expressions we encounter two types of divergent terms: one with
negative power of $\delta$ and one with an integral of $\sin^{-2}(\pi x/L)$,
which diverges at $x = 0$ and $x = L$. These terms, however, are free of the
source parameter $\phi$ and hence do not change any observables. The
$\phi$-dependent terms are finite and, after integrating over $\bm{\delta}$ in
Eq.\ (\ref{Omegaa}), yield the simple result:\cite{footnote-circular}
\begin{equation}
 \Omega_a
  = \mathrm{const} + (\alpha_0 - \alpha_z)\, \frac{W \phi^2}{4 \pi L}.
 \label{Omegaa_res}
\end{equation}
It provides a linear (in $\alpha_\mu$) correction to the free energy of the
clean sample, Eq.\ (\ref{Omega0}), merely changing the overall prefactor
(conductance) but preserving the quadratic dependence on $\phi$ and hence the
form of the Dorokhov distribution. Thus the linear disorder correction does not
destroy the pseudodiffusive character of transport in graphene at the Dirac
point.

\subsection{Second-order corrections}

Since the lowest disorder correction (\ref{Omegaa_res}) preserves the form of
the Dorokhov distribution, we proceed with higher order corrections. Our aim is
to find a deviation from the pseudodiffusive transport. The second-order
correction to the free energy is due to the diagrams Fig.\ \ref{Fig:diagrams}b
and \ref{Fig:diagrams}c. The diagram with parallel impurity lines (Fig.\
\ref{Fig:diagrams}b) yields
\begin{align}
 \Omega_b
  = &\int d\bm{\delta}\, d\bm{\delta}' \sum_{\mu,\nu} w_\mu(|\bm{\delta}|)
    w_\mu(|\bm{\delta}'|) \Omega_b^{(\mu\nu)}, \\
 \Omega_b^{(\mu\nu)}
  = 2\pi^2 &\int d\mathbf{r}\, d\mathbf{r}' \mathop{\mathrm{Tr}} \Big[
      \sigma_\mu G(\mathbf{r}, \mathbf{r} + \bm{\delta})
      \sigma_\mu G(\mathbf{r} + \bm{\delta}, \mathbf{r}') \notag\\
      &\times \sigma_\nu G(\mathbf{r}', \mathbf{r}' + \bm{\delta}')
      \sigma_\nu G(\mathbf{r}' + \bm{\delta}', \mathbf{r})
    \Big].
\end{align}
Using the Green function from Eq.\ (\ref{G0}), we expand the correction to the
free energy in powers of $\bm{\delta}$ and $\bm{\delta}'$. Then we drop all
$\phi$-independent terms and average with respect to the directions of
$\bm{\delta}$ and $\bm{\delta}'$. The following four contributions to the
free energy are non zero:
\begin{multline}
 \Omega_b^{(00)}
  = \Omega_b^{(zz)}
  = -\Omega_b^{(0z)}
  = -\Omega_b^{(z0)} \\
  = \frac{W \phi^2}{64 L^4} \int_0^L dx\, dx' \int_{-\infty}^\infty dy \\
    \times\left[
        \frac{1}{\sin^2 \frac{\pi (x + x' + iy)}{2L}}
        +\frac{1}{\sin^2 \frac{\pi (x - x' + iy)}{2L}}
        +\mathrm{c.c.}
    \right].
\end{multline}
After integrating with respect to $x$ and $x'$ the above expression vanishes.
Thus we conclude that the diagram Fig.\ \ref{Fig:diagrams}b gives no
contribution to the free energy,
\begin{equation}
 \Omega_b
  = 0.
\end{equation}

Let us now consider the diagram Fig.\ \ref{Fig:diagrams}c with crossed impurity
lines. This diagram contains no Green functions at coincident points and hence
does not require regularization. We can replace disorder correlation functions
$w_\mu$ by equivalent delta-functions and obtain
\begin{gather}
 \Omega_c
  = \sum_{\mu\nu} \alpha_\mu \alpha_\nu \Omega_c^{(\mu\nu)}, \label{Omegacs}\\
 \Omega_c^{(\mu\nu)}
  = \pi^2 \int d\mathbf{r}\, d\mathbf{r}'
    \mathop{\mathrm{Tr}} \Big[
      \sigma_\mu \check G(\mathbf{r}, \mathbf{r}')
      \sigma_\nu \check G(\mathbf{r}', \mathbf{r})
    \Big]^2.
\end{gather}
With the Green function (\ref{G0}) we find the following contribution to the
sum in Eq.\ (\ref{Omegacs}):
\begin{align}
 \Omega_c^{(00)}
  &= \Omega_c^{(0z)}
  = \Omega_c^{(z0)} \notag \\
  &= \frac{\pi^2 W}{64 L^4} \int\limits_0^L dx\, dx'
    \int\limits_{-\infty}^\infty dy \cosh \frac{2 \phi y}{L} \\
    &\times\left(
      \frac{1}{\big| \sin \frac{\pi (x + x' + iy)}{2L} \big|^2}
      -\frac{1}{\big| \sin \frac{\pi (x - x' + iy)}{2L} \big|^2}
    \right)^2, \notag \\
 \Omega_c^{(zz)}
  &= \Omega_c^{(00)} + \frac{\pi^2 W}{8 L^4} \int\limits_0^L dx\, dx'
    \int\limits_{-\infty}^\infty dy \cosh \frac{2 \phi y}{L} \notag \\
    &\times \left|
      \sin \frac{\pi (x + x' + iy)}{2L}
      \sin \frac{\pi (x - x' + iy)}{2L}
    \right|^{-2}.
\end{align}
Two-dimensional integrals with respect to $x$ and $x'$ are straightforward
due to periodicity of the integrand. As a result, the free energy is expressed
as a single $y$ integral:
\begin{multline}
 \Omega_c
  = \frac{\pi^2 W}{8 L^2} \int\limits_{-\infty}^\infty dy\,
    \frac{\cosh(2\phi y/L)}{\sinh^2 (\pi y/L)} \bigg[
      (\alpha_0 + \alpha_z)^2 \coth \left| \frac{\pi y}{L} \right|\\
      -(\alpha_0 + 3 \alpha_z) (\alpha_0 - \alpha_z)
    \bigg].
 \label{Omegac}
\end{multline}
This integral diverges at $y = 0$. Expanding near this point, we find that the
integrand behaves as $(L/|y|)^3 + 2 \phi^2 L / |y|$. The most singular part is
$\phi$-independent and hence unobservable. Integral of the second term diverges
logarithmically and multiplies $\phi^2$. This gives a logarithmic correction
to the conductance of the system preserving the pseudodiffusive form of the
transmission distribution. Let us cut off the logarithmic integral at some
ultraviolet scale $y = a$ that is the smallest scale where the massless Dirac
model with Gaussian white-noise disorder applies, e.g. the scale of the disorder
correlation length or lattice spacing in graphene. The upper cut-off is already
embedded in the integrand of Eq.\ (\ref{Omegac}): the small $y$ expansion is
valid for $y \lesssim L$. Thus we can isolate the divergent part of the integral
(\ref{Omegac}) and the remaining $\Omega_c$ correction, which has a
non-trivial dependence on $\phi$.
\begin{multline}
 \Omega_c
  = \frac{W \phi^2}{4\pi L} \Big\{
      (\alpha_0 + \alpha_z)^2 \big[ 2 \ln(L/a) + \omega_1(\phi) \big] \\
      +(\alpha_0 + 3 \alpha_z) (\alpha_0 - \alpha_z) \omega_2(\phi)
    \Big\}.
 \label{Omegac_res}
\end{multline}
Since the logarithmic term in the free energy contains an ultraviolet parameter
$a$ defined up to a model-dependent constant, the functions $\omega_{1,2}(\phi)$
are fixed up to an arbitrary constant. With this accuracy, we find
\begin{align}
 \omega_1(\phi)
  &= \frac{\pi^3}{2L\phi^2} \!\int\limits_{-\infty}^\infty\! dy
    \frac{\cosh\frac{\pi y}{L}}{\sinh^3\big|\frac{\pi y}{L}\big|} \left(
      \cosh\frac{2\phi y}{L} - 1 - \frac{2 \phi^2 y^2}{L^2}
    \right) \notag\\
  &= \mathrm{const} - \psi(\phi/\pi) - \psi(-\phi/\pi), \label{omega1} \\
 \omega_2(\phi)
  &= -\frac{\pi^3}{2L\phi^2} \int\limits_{-\infty}^\infty dy
    \frac{\cosh(2 \phi y/L) - 1}{\sinh^2(\pi y/L)} \notag\\
  &= \mathrm{const} + \pi^2 \frac{\phi \cot \phi - 1}{\phi^2}. \label{omega2}
\end{align}

The logarithmic correction in Eq.\ (\ref{Omegac_res}) can be included into an
effective $L$-dependence of the disorder strength parameters $\alpha_\mu$ by
renormalization group (RG) methods. The model of two-dimensional massless
Dirac fermions subject to Gaussian disorder and its logarithmic renormalization
appeared in various contexts. In particular, disorder renormalization in
graphene was considered in Refs.\ \onlinecite{Aleiner06, OurPRB, Schuessler09}.
One-loop RG equations for effective disorder couplings as functions of a running
scale $\Lambda$ are
\begin{subequations}
\begin{align}
 \frac{\partial\alpha_0}{\partial \ln \Lambda}
  &= 2 (\alpha_0 +\alpha_z) (\alpha_0 + \alpha_x + \alpha_y), \label{RGa0}\\
 \frac{\partial\alpha_x}{\partial \ln \Lambda}
  &= \frac{\partial\alpha_y}{\partial \ln \Lambda}
  = 2 \alpha_0 \alpha_z, \label{RGaperp}\\
 \frac{\partial\alpha_z}{\partial \ln \Lambda}
  &= 2 (\alpha_0 + \alpha_z) (-\alpha_z + \alpha_x + \alpha_y).  \label{RGaz}
\end{align}
\label{RG}
\end{subequations}
Parameters defined in Eq.\ (\ref{amu}) serve as initial conditions for the
RG equations at an ultraviolet scale $a$. Integrating Eqs.\ (\ref{RG}) up to the
largest scale that is the system size $L$, we obtain effective disorder
couplings $\tilde\alpha_\mu = \alpha_\mu(L)$ and automatically take into account
all leading logarithmic contributions like the one in Eq.\ (\ref{Omegac_res}).
This allows us to replace the disorder parameters in the free energy by their
renormalized values and drop the logarithm from Eq.\ (\ref{Omegac_res}).
Collecting in this way the contributions (\ref{Omega0}), (\ref{Omegaa_res}), and
(\ref{Omegac_res}), we obtain the final expression for the free energy up to
the second order in renormalized disorder parameters,
\begin{multline}
 \Omega
  = \frac{W \phi^2}{4\pi L} \Big[
      1 + \tilde\alpha_0 - \tilde\alpha_z
      +(\tilde\alpha_0 + \tilde\alpha_z)^2 \omega_1(\phi) \\
      +(\tilde\alpha_0 + 3 \tilde\alpha_z) (\tilde\alpha_0 - \tilde\alpha_z)
        \omega_2(\phi)
    \Big].
 \label{Omega_res}
\end{multline}

Thus we have established a deviation from pseudodiffusive transport regime
($\Omega \sim \phi^2$) in the second order in disorder strength.

\subsection{Corrections to the distribution function}

Let us now derive a correction to the Dorokhov distribution function of
transmission probabilities. In the $\lambda$ representation, the distribution
function is given by Eq.\ (\ref{PfromO}). Using the result (\ref{Omega_res}), we
obtain
\begin{multline}
 P(\lambda)
  = \frac{W}{\pi L} \Big[
      1 + \tilde\alpha_0 - \tilde\alpha_z
      +(\tilde\alpha_0 + \tilde\alpha_z)^2 p_1(\lambda) \\
      +(\tilde\alpha_0 + 3 \tilde\alpha_z) (\tilde\alpha_0 - \tilde\alpha_z)
        p_2(\lambda)
    \Big].
 \label{P_res}
\end{multline}
Similarly to $\omega_{1,2}$, the functions $p_{1,2}(\lambda)$ are defined up to
a model-dependent constant. From Eqs.\ (\ref{omega1}) and (\ref{omega2}) we
obtain
\begin{subequations}
\begin{align}
 p_1(\lambda)
  &= \mathrm{const}
    -2 \mathop{\mathrm{Re}} \frac{\partial}{\partial\lambda} \left[
      \lambda\; \psi\left(\frac{2i\lambda}{\pi}\right)
    \right], \\
 p_2(\lambda)
  &= \mathrm{const}
    +\frac{\pi^2}{2 \sinh^2(2\lambda)}.
\end{align}
\label{p12}
\end{subequations}

\begin{figure}
 \centerline{\includegraphics[width=0.9\linewidth]{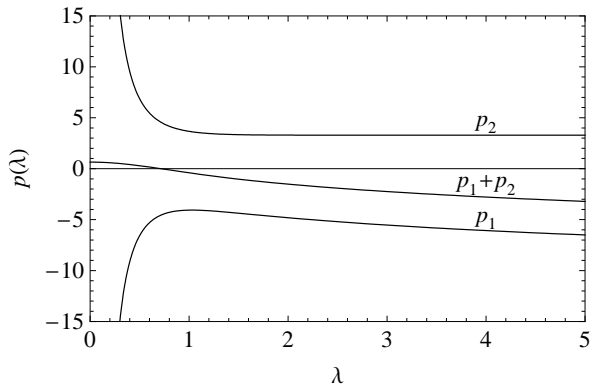}}
 \caption{Functions $p_1$ and $p_2$ entering the disorder correction to the
distribution of transmission probabilities (\protect\ref{P_res}). In the case
of random scalar potential ($\alpha_0$), the distribution is determined by the
sum $p_1 + p_2$ only.}
 \label{fig:p1p2p12}
\end{figure}

The functions $p_1$ and $p_2$ are shown in Fig.\ \ref{fig:p1p2p12}. (When
the only disorder is $\alpha_0$, correction to the distribution function is
given by the sum $p_1 + p_2$ also shown in the figure.) The functions $p_1$ and
$p_2$ cannot be used for direct calculation of transmission moments due to their
divergence at $\lambda = 0$. This divergence signifies the break down of
perturbative expansion in small values of disorder couplings close to $\lambda =
0$ (that is $T = 1$). Comparing disorder correction with the distribution in the
clean sample, we conclude that the result (\ref{P_res}) is valid provided
$\lambda \gg \tilde\alpha$.

The deviation from pseudodiffusive transport regime can be experimentally
demonstrated as a correction to the Fano factor $F = 1/3$ characteristic to the
diffusive systems. Divergence of the functions $p_{1,2}$ at $\lambda = 1$
prevents us from calculating transmission moments from the distribution function
(\ref{P_res}). However, we can obtain transport characteristics from the free
energy (\ref{Omega_res}) instead. With the help of Eq.\ (\ref{Fdef}), we
find the Fano factor up to quadratic terms in the renormalized disorder
strength,
\begin{multline}
 F
  = \frac{1}{3}
    -\frac{16 \zeta(3)}{\pi^2} (\tilde\alpha_0 + \tilde\alpha_z)^2
    +\frac{8 \pi^2}{45} (\tilde\alpha_0 + 3 \tilde\alpha_z)
      (\tilde\alpha_0 - \tilde\alpha_z) \\
  \approx \frac{1}{3}
    -0.194\, \tilde\alpha_0^2
    -0.388\, \tilde\alpha_0 \tilde\alpha_z
    -7.212\, \tilde\alpha_z^2.
 \label{Fball}
\end{multline}
Remarkably, any weak disorder, irrespective of its matrix structure, suppresses
the Fano factor. [Note that the energy correction (\ref{GFsmallE}) is also
negative.] The correction to the Fano factor increases with increasing sample
length $L$ due to renormalization (\ref{RG}). At some length $l$, referred to
as the mean free path, one of the renormalized disorder couplings reach a value
of order unity and the perturbative RG treatment breaks down. This signifies the
crossover from ballistic to diffusive transport regime. Disorder correction to
the Fano factor becomes strong in this crossover region. To go beyond the mean
free path scale we resort to other methods designed for diffusive systems.

\section{Disordered graphene: diffusive limit}
\label{sec:dl}

When the system size exceeds the mean free path, the sample exhibits diffusive
electron transport. On a semiclassical level, the system can be characterized by
its conductivity per square in this limit. At the ballistics-diffusion
crossover the conductivity of graphene is close to the quantum value $e^2/h$.
This signifies strong interference corrections to transport characteristics
making semiclassical picture inadequate. These quantum effects lead to one
of the four possible scenarios depending on the symmetry of disorder:

\begin{enumerate}
\item[(i)] If the only disorder is random potential ($\alpha_0$), the system
possesses time inversion symmetry $H = \sigma_2 H^T \sigma_2$ and falls into
symplectic symmetry class AII\cite{AltlandZirnbauer, MirlinEvers}. Quantum
corrections to the conductivity are positive, leading to good metallic
properties (large dimensionless conductivity) at large scales.

\item[(ii)] In the case of random vector potential $\alpha_{x,y}$, the only
symmetry of the problem is chirality, $H = -\sigma_3 H \sigma_3$, signifying the
chiral unitary symmetry class AIII. Such disorder produces no corrections to the
conductivity to all orders and can be effectively gauged out at zero energy
\cite{Schuessler09}. From the point of view of its transport properties, the
system remains effectively clean and ballistic at all scales.

\item[(iii)] If the only disorder is random mass ($\alpha_z$), the Hamiltonian
has a Bogolyubov -- de Gennes symmetry $H = \sigma_1 H^T \sigma_1$
characteristic for the symmetry class D. Upon renormalization (\ref{RG}) the
disorder coupling gets smaller and the system becomes effectively clean. This
means the absence of the mean free path scale and hence of the diffusive
transport regime.

\item[(iv)] In the generic case, when more than one disorder type is present and
all symmetries are broken, the symmetry class is unitary (A) and transport
properties are the same as at the critical point of the quantum Hall transition.
\end{enumerate}

We will concentrate on the first case (random potential) when the system
eventually acquires a large parameter -- dimensionless conductivity -- and can
be quantitatively described by the proper effective field theory -- sigma model
of the symplectic symmetry class. Our consideration in this part of the
paper is closely related to that of Ref.\ \onlinecite{Nazarov95}.

Derivation of the sigma model with the source fields $z$ from Eq.\
(\ref{matrixG}) is sketched in Appendix \ref{AppSigma}. The symplectic sigma
model operates with the matrix field $Q$ of the size $4N \times 4N$, where $N$
is the number of replicas. Apart from replica space, matrix $Q$ has
retarded-advanced (RA) and particle-hole (PH) structures. The former is similar
to the matrix Green function while the latter is introduced in order to account
for time-reversal symmetry of the problem. We will denote Pauli matrices in RA
space by $\Lambda_{x,y,z}$. Two constraints are imposed on $Q$, namely, $Q^2 =
1$ and $Q = Q^T$. This yields the target space $Q \in O(4N)/O(2N) \times O(2N)$
characteristic for symplectic class systems. The sigma-model action
is\cite{footnote-top-term}
\begin{equation}
 S[Q]
  = \frac{\sigma}{16} \int d\mathbf{r} \mathop{\mathrm{Tr}} (\nabla Q)^2.
 \label{sigma}
\end{equation}
Here $\sigma$ is the dimensionless (in units $e^2/h$) conductivity of the
two-dimensional disordered system. The source field is incorporated into
boundary conditions,
\begin{equation}
 Q|_{x = 0}
  = \Lambda_z,
 \qquad
 Q|_{x = L}
  = \Lambda_z \cos\phi + \Lambda_x \sin\phi.
 \label{boundary}
\end{equation}
The free energy of the system in the source field $\phi$ is expressed through
the $N \to 0$ limit of the sigma-model partition function as
\begin{equation}
 \Omega
  = \lim_{N \to 0} \frac{1}{N} \left( 1 - \int DQ e^{-S[Q]} \right).
 \label{Omega_sigma}
\end{equation}

\begin{figure}
 \centerline{\includegraphics[height=0.3\linewidth]{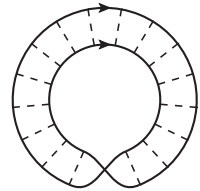}}
 \caption{Cooperon correction to the free energy in the diffusive limit.}
\label{fig:Cooperon}
\end{figure}

In a good metallic sample with $\sigma \gg 1$, the $Q$ integral in Eq.\
(\ref{Omega_sigma}) can be evaluated within the saddle point approximation.
The action (\ref{sigma}) is minimized by the following configuration of the
field $Q$:
\begin{equation}
 Q_0
  = U^{-1} \Lambda_z U,
 \qquad
 U
  = \exp\left( i \Lambda_y \frac{\phi x}{2L} \right).
\end{equation}
Replacing the integral in Eq.\ (\ref{Omega_sigma}) with the value of the
integrand at the saddle point, we obtain the semiclassical result for the full
counting statistics,
\begin{equation}
 \Omega_0
  = \lim_{N \to 0} \frac{S[Q_0]}{N}
  = \frac{W \sigma \phi^2}{4 L}.
\end{equation}
This yields the Dorokhov distribution of transmission probabilities in diffusive
two-dimensional system.\cite{Dorokhov83} In order to find corrections to this
result, we take into account fluctuations of the field $Q$ near its saddle-point
value $Q_0$. This is equivalent to the calculation of a Cooperon loop, Fig.\
\ref{fig:Cooperon}, carried out in Ref.\ \onlinecite{Nazarov95}.

Small fluctuations of $Q$ near the saddle point $Q_0$ are parametrized by the
matrix $B$ as (we write expressions involving $B$ up to the second order)
\begin{equation}
 Q
  = U^{-1} \Lambda_z \left(
      1 + B + \frac{B^2}{2}
    \right) U,
 \qquad
 B
  = \begin{pmatrix}
      0 & b \\
      -b^T & 0
    \end{pmatrix}.
\end{equation}
This parametrization of $Q$ automatically fulfils the conditions $Q^2 = 1$ and
$Q = Q^T$. The sigma-model action expanded up to the second order in $B$ takes
the form
\begin{equation}
 S[Q]
  = S[Q_0] - \frac{\sigma}{16} \int d\mathbf{r} \mathop{\mathrm{Tr}} \left[
      (\nabla B)^2 - \frac{\phi^2}{4L^2} \{\Lambda_x, B\}^2
    \right].
 \label{sigma_B}
\end{equation}
Curly braces denote anticommutator. Let us separate $B$ into the parts commuting
and anticommuting with $\Lambda_x$. These two parts do not couple to each other
in the quadratic action (\ref{sigma_B}) and only the former one couples to the
source parameter $\phi$. Thus we can constraint the matrix $B$ by requiring its
commutativity with $\Lambda_x$. In terms of $b$ this yields $b = -b^T$ and the
action becomes
\begin{equation}
 S[Q]
  = S[Q_0] + \frac{\sigma}{8} \int d\mathbf{r} \mathop{\mathrm{Tr}} \left[
      \nabla b \nabla b^T - \frac{\phi^2}{L^2}\; b b^T
    \right].
 \label{sigma_b}
\end{equation}
This quadratic form is diagonalized in momentum representation. Component of
momentum perpendicular to the leads takes quantized values $\pi n/L$ with
positive integer $n$ due to geometrical restrictions [boundary conditions
(\ref{boundary}) fix $b = 0$ at the interfaces with metallic leads]. Momentum
parallel to the leads is continuous and unrestricted. For each value of
the momentum there are $N(2N-1)$ independent matrix elements in $b$. Calculating
the Gaussian integral in Eq.\ (\ref{Omega_sigma}) we obtain the free energy
\begin{multline}
 \Omega
  = \Omega_0 - \frac{W}{2} \sum_{n=1}^\infty \int \frac{dq_y}{2\pi} \ln \big(
      \pi^2 n^2 + q^2 L^2 - \phi^2
    \big) \\
  = \frac{W}{2L} \left[
      \frac{\sigma \phi^2}{2} - \sum_{n=1}^\infty \sqrt{\pi^2 n^2 - \phi^2}
    \right].
 \label{Omega_dif}
\end{multline}

In the result (\ref{Omega_dif}), the sum diverges at large $n$. The situation is
similar to what we have encountered in the ballistic regime. Expanding the sum
in powers of $\phi$, we see that the most divergent term is $\phi$-independent,
while the next term multiplies $\phi^2$ and diverges logarithmically. This is
nothing but the weak antilocalization correction. It renormalizes the
conductivity but does not deform the full counting statistics. Logarithmically
divergent sum is cut at $n \sim L/l$ where $l$ is the mean free path. At larger
values of $n$ the diffusive approximation (gradient expansion in the sigma
model) breaks down. In terms of renormalized conductivity, the free energy reads
\begin{gather}
 \Omega
  = \frac{W}{2L} \left[
      \frac{\tilde \sigma \phi^2}{2} - \sum_{n=1}^\infty \left(
        \sqrt{\pi^2 n^2 - \phi^2} - \pi n + \frac{\phi^2}{2\pi n}
      \right)
    \right], \label{Omega_dif_renorm} \\
 \tilde\sigma
  = \sigma + \frac{1}{\pi} \ln \frac{L}{l}
  \approx \frac{1}{\pi} \ln \frac{L}{l}.
 \label{tsigma}
\end{gather}
The bare value of conductivity, $\sigma$, is of order one and hence negligible
in comparison with the large renormalizing logarithm. The sum over $n$ in Eq.\
(\ref{Omega_dif_renorm}) is convergent and provides the deviation from
semiclassical Dorokhov statistics of transmission probabilities.

In fact, a more rigorous procedure is to perform first a renormalization of the
sigma model from the mean free path scale $l$ to the scale $\sim L$. Then the
free energy can be calculated perturbatively. It turns out, however, that this
yields a result identical to the one obtained above within the perturbative
analysis at the scale $l$. Indeed, the RG equation $d\sigma/d\ln\Lambda =
1/\pi$ will lead exactly to the renormalization of conductivity $\sigma \mapsto
\tilde\sigma$, see Eq.\ (\ref{tsigma}). The consequent evaluation of the
perturbative contribution to $\Omega$ yields Eq.\ (\ref{Omega_dif}) with
$\sigma$ replaced by $\tilde\sigma$ and the sum restricted to a finite
(independent of $L$) number of terms. In other words, the renormalization shifts
the logarithmical contribution to $\sigma$ from the second to the first term in
square brackets in Eq.\ (\ref{Omega_dif}).

\begin{figure}
 \centerline{\includegraphics[width=0.9\linewidth]{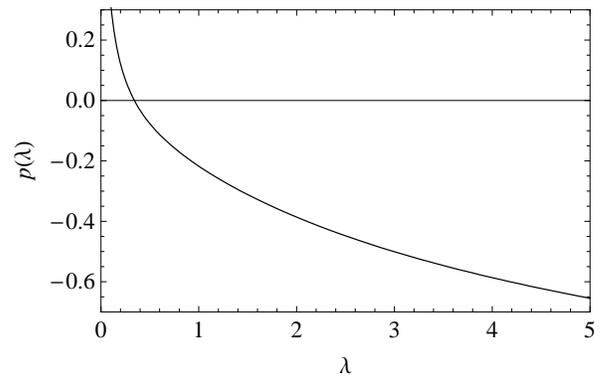}}
 \caption{Correction to the distribution of transmission eigenvalues in the
diffusive limit.}
\label{fig:p}
\end{figure}

Let us derive the distribution function $P(\lambda)$ from the free energy
(\ref{Omega_dif_renorm}). Applying Eq.\ (\ref{PfromO}), we obtain the result in
the form
\begin{gather}
 P(\lambda)
  = \frac{W}{L} \big[ \tilde\sigma + p(\lambda) \big],
 \label{Pdif}\\
 p(\lambda)
  = \frac{1}{\pi} \sum_{n = 1}^\infty \left[ \mathop{\mathrm{Re}}
      \frac{\pi + 2 i \lambda}{\sqrt{\pi^2 n^2 - (\pi + 2 i \lambda)^2}}
      -\frac{1}{n}
    \right].
 \label{p}
\end{gather}
At small values of $\lambda$, the sum in Eq.\ (\ref{p}) is determined by the
term with $n = 1$. In the opposite limit, the sum can be estimated by the
corresponding integral with the help of Euler-Maclaurin formula. Thus we
obtain the asymptotic expressions
\begin{equation}
 p(\lambda)
  = \begin{cases}
      \sqrt{\dfrac{1}{8\pi\lambda}}, & \lambda \ll 1, \\[10pt]
      -\dfrac{1}{\pi}\ln\lambda, & \lambda \gg 1.
    \end{cases}
\end{equation}
The function $p(\lambda)$ is shown in Fig.\ \ref{fig:p}. It is qualitatively
similar to the numerical result of Ref.\ \onlinecite{San-Jose07}.

Deviation from the semiclassical transport can be demonstrated by the
correction to the Fano factor. With the help of Eq.\ (\ref{Fdef}), we obtain
\begin{equation}
 F
  = \frac{1}{3} - \frac{2 \zeta(3)}{\pi^3 \tilde\sigma}
  = \frac{1}{3} - \frac{0.244}{\ln(L/l)}.
 \label{Fdif}
\end{equation}
A similar correction to the Fano factor was found numerically in Ref.\
\onlinecite{Tworzydlo08}. We compare the numerical results with Eq.\
(\ref{Fdif}) below.

\begin{figure}
 \centerline{\includegraphics[width=0.9\linewidth]{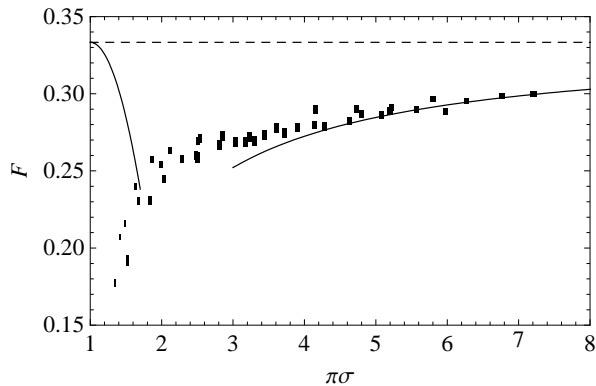}}
 \caption{Fano factor as a function of conductivity. Solid lines show ballistic
and diffusive results (\protect\ref{Fsigma}). Dashed line corresponds to the
asymptotic value $F = 1/3$. Solid symbols are numerical results from Ref.\
\protect\onlinecite{Tworzydlo08}, the size of rectangles corresponds to the
error estimate.}
\label{fig:F}
\end{figure}

In the case of weak scalar disorder (described by the coupling $\alpha_0$), the
system undergoes a continuous crossover from ballistic to diffusive transport
regime as the size $L$ grows. In both limiting cases, we encounter nearly
Dorokhov distribution of transmission probabilities with small corrections,
Eqs.\ (\ref{Fball}) and (\ref{Fdif}), on both sides of the crossover. In the
ballistic limit, we can formally introduce a dimensionless conductivity as
$\sigma = (L/W) G/(e^2/h)$. Then the corrections to the Fano factor are
expressed in terms of the conductivity
\begin{equation}
 F
  = \frac{1}{3} - \begin{cases}
      \left(
        \dfrac{16\zeta(3)}{\pi^2} - \dfrac{8\pi^2}{45}
      \right)(\pi\sigma - 1)^2, & \pi \sigma - 1 \ll 1, \\[10pt]
      \dfrac{2\zeta(3)}{\pi^3 \sigma}, & \sigma \gg 1.
    \end{cases}
 \label{Fsigma}
\end{equation}
This Fano factor as a function of conductivity is shown in Fig.\ \ref{fig:F}
together with numerical results from Ref.\ \onlinecite{Tworzydlo08}. In 
the numerical simulations, a single valley of graphene was modeled using a
finite-difference approach. By construction, disorder in Ref.\
\onlinecite{Tworzydlo08} has the symmetry of scalar potential which does not
mix the valleys. It is this symmetry (class AII) which is considered in the
present section. Our results perfectly agree with the numerics in the
diffusive limit (see Fig.\ \ref{fig:F}) in the range $\pi\sigma \gtrsim 3$. On
the ballistic side, the deviation is due
to the non-universality of the ballistic transport. Specifically, the function
$F(\sigma)$ depends crucially on the microscopic details of disorder. In the
numerical analysis of Ref.\ \onlinecite{Tworzydlo08}, the model with strong
scatterers was used, while in the present paper we adopt the model of weak
Gaussian white-noise disorder. For theoretical predictions on electron transport
in the presence of strong scatterers see Ref.\ \onlinecite{Titov10}.

An earlier numerical study of Ref.\ \onlinecite{San-Jose07}, based on the
transfer-matrix description of the Dirac problem, reported the value of the Fano
factor in the range $0.29 \div 0.30$ (for different samples) with the
conductivity, $\pi\sigma$, of the largest systems varying from $6$ to $10$. This
is consistent with our predictions for the diffusive transport regime (see Fig.\
\ref{fig:F}). The behavior of $F$ in the ballistic regime is different due to
the reasons described above (strong vs. weak disorder). A non-monotonous
dependence $F(\sigma)$ at the Dirac point was also observed in Ref.\
\onlinecite{Lewenkopf08}.

The Fano factor is $1/3$ both in the clean and strongly disordered limits. In
the crossover from ballistics to diffusion, the Fano factor strongly deviates
from this universal value signifying the breakdown of the (pseudo-) diffusive
description characterized by Dorokhov distribution of transmission
probabilities.

\section{Summary}
\label{sec:sum}

We have studied the full counting statistics of the charge transport through an
undoped graphene sheet in the presence of weak and smooth (not mixing valleys)
disorder. We have identified deviations from the Dorokhov distribution of
transmission probabilities both in ballistic [Eq.\ (\ref{P_res}), (\ref{p12})]
and diffusive [Eqs.\ (\ref{Pdif}), (\ref{p})] regimes. In the former case,
corrections are model-dependent while in the latter case only the symmetry of
disorder matters. We have considered Gaussian white-noise disorder in the
ballistic regime and potential disorder (symplectic symmetry class) in diffusive
limit. Deviation from (pseudo-) diffusive transport always results in a negative
correction to the Fano factor, $F < 1/3$. Our results are in good agreement with
recent numerical simulations of electron transport in disordered graphene, see
Fig.\ \ref{fig:F}.

\begin{acknowledgments}
We are grateful to R.\ Danneau, P.\ San-Jose, and M.\ Titov for stimulating
discussions and to C.\ Groth for providing us with the numerical data of Ref.\
\onlinecite{Tworzydlo08}. The work was supported by Rosnauka grant
02.740.11.5072 and by the EUROHORCS/ESF EURYI Award scheme (I.V.G.).
\end{acknowledgments}

\onecolumngrid
\appendix

\section{Matrix Green function}
\label{AppG}

The full counting statistics of the electron transport is conveniently
expressed in terms of the matrix Green function \cite{Nazarov94} in the
external counting field $z = \sin^2(\phi/2)$, Eq.\ (\ref{matrixG}). For the
clean graphene sample attached to perfect metallic leads, Fig.\
\ref{fig:sample}, this Green function satisfies the following equation:
\begin{equation}
 \label{keldysh}
 \begin{pmatrix}
   \mu(x) - \bm{\sigma} \mathbf{p} + i0 & -\sigma_x \sqrt{z} \delta(x) \\
   -\sigma_x \sqrt{z} \delta(x-L) & \mu(x) - \bm{\sigma} \mathbf{p} - i0
 \end{pmatrix} \check{G}_0 (\mathbf{r}, \mathbf{r}')
  = \delta(\mathbf{r} - \mathbf{r}'),
 \qquad\qquad
 \mu(x)
  = \begin{cases}
      0, & 0 < x < L, \\
      +\infty, & \text{$x < 0$ or $x > L$}.
    \end{cases}
\end{equation}
Since the operator in the left-hand side of the above equation commutes with the
$y$ component of the momentum, we will first calculate the Green function in the
mixed coordinate-momentum representation, $\check G_p (x, x')$. Inside the
sample this function satisfies
\begin{equation}
 \label{eqGreen}
 \left[
   i \sigma_x \frac{\partial}{\partial x} - \sigma_y p
 \right] \check G_p (x, x')
  = \delta(x - x').
\end{equation}

We will look for a general solution of this equation in the form
\begin{equation}
 \check G_p (x, x')
  = e^{\sigma_z p (x - L/2)} M e^{\sigma_z p (x' - L/2)},
 \qquad\qquad
 M
  = \begin{cases}
      M_<, & x < x', \\
      M_>, & x > x'.
    \end{cases}
\end{equation}
The chemical potential profile together with the infinitesimal terms $\pm i0$
in Eq.\ (\ref{keldysh}) defines the boundary conditions for the Green function.
The counting field $z$ can also be incorporated into the boundary conditions. In
terms of $M_\lessgtr$ we thus obtain
\begin{equation}
 \label{boundaryG}
 \begin{pmatrix}
   1 & 1 & i\sqrt{z} & i\sqrt{z} \\
   0 & 0 & 1 & -1
 \end{pmatrix} e^{-\sigma_z p L/2} M_<
  = 0,
 \qquad
 \begin{pmatrix}
   1 & -1 & 0 & 0 \\
   -i\sqrt{z} & -i\sqrt{z} & 1 & 1
 \end{pmatrix} e^{\sigma_z p L/2} M_>
  = 0.
\end{equation}
Delta function in the right-hand side of Eq.\ (\ref{eqGreen}) yields a jump of
the Green function at $x = x'$ which provides the relation
\begin{equation}
 \label{jump}
 M_> - M_<
  = -i \sigma_x.
\end{equation}
The matrices $M_\lessgtr$, and hence the Green function, are completely
determined by Eqs.\ (\ref{boundaryG}) and (\ref{jump}),
\begin{equation}
 M_\lessgtr
  = \frac{-i}{2(\cosh^2 pL - z)}\begin{pmatrix}
        \cosh pL & z - \dfrac{\sinh 2pL}{2} & i\sqrt{z} e^{-pL} & i\sqrt{z}\\
        z + \dfrac{\sinh 2pL}{2} & \cosh pL & i\sqrt{z} & i\sqrt{z} e^{pL} \\
        i\sqrt{z} e^{pL} & i\sqrt{z} & -\cosh pL & -z - \dfrac{\sinh 2pL}{2} \\
        i\sqrt{z} & i\sqrt{z} e^{-pL} & -z + \dfrac{\sinh 2pL}{2} & -\cosh pL
    \end{pmatrix} \pm \frac{i\sigma_x}{2}.
\end{equation}

Fourier transform in $p$ yields the Green function in the full coordinate
representation. To facilitate further calculations, we decompose this Green
function into the following product of matrices:
\begin{gather}
 \check G_0 (x, x'; y)
  = \frac{1}{4L}\, \check U(x) \check \Lambda \begin{pmatrix}
      i \cosh\dfrac{\phi y}{2L} & \sinh\dfrac{\phi y}{2L} \\[8pt]
      \sinh\dfrac{\phi y}{2L} & -i \cosh\dfrac{\phi y}{2L}
    \end{pmatrix}_{RA} \begin{pmatrix}
      \dfrac{1}{\sin \frac{\pi}{2L} (x + x' + iy)} &
      \dfrac{1}{\sin \frac{\pi}{2L} (x - x' + iy)} \\[8pt]
      \dfrac{1}{\sin \frac{\pi}{2L} (x - x' - iy)} &
      \dfrac{1}{\sin \frac{\pi}{2L} (x + x' - iy)}
    \end{pmatrix}_\sigma \check \Lambda \check U^{-1}(x'), \label{G0} \\
 \check \Lambda
  = \begin{pmatrix}
      \sigma_z & 0 \\
      0 & 1
    \end{pmatrix}_{RA},
 \qquad\qquad
 \check U(x)
  = \begin{pmatrix}
      \sin\dfrac{\phi (L - x)}{2L} & \cos\dfrac{\phi (L - x)}{2L} \\[8pt]
      i\cos\dfrac{\phi x}{2L} & i \sin\dfrac{\phi x}{2L}
    \end{pmatrix}_{RA}.
\end{gather}
Here we have used the source angle $\phi$ defined by $z = \sin^2(\phi/2)$. The
matrices $\check U(x)$ and $\check U^{-1}(x')$ operate in the retarded-advanced
space only and hence commute with any disorder operators placed between the
Green functions. As a result, factors $\check U$ and $\check U^{-1}$ drop from
expressions for any closed diagrams. The matrices $\check \Lambda$ in the above
equation allow us to decompose the Green function into a direct product of the
two operators acting in the RA space and in the sublattice space.

\section{Energy correction to the full counting statistics}
\label{AppE}

In this appendix we evaluate the diagram Fig.\ \ref{fig:eloop} for the lowest
energy correction to $\Omega(\phi)$. Substituting Green function (\ref{G0}) into
Eq.\ (\ref{Oe}) and performing rescaling of integration variables we obtain
\begin{equation}
 \Omega_\epsilon
  = \frac{WL\epsilon^2}{4}
    \int\limits_0^1 dx\, dx' \int\limits_{-\infty}^\infty dy
    \cosh(\phi y)\left[
      \frac{1}{\cosh(\pi y) - \cos \pi(x + x')}
      -\frac{1}{\cosh(\pi y) - \cos \pi(x - x')}
    \right].
\end{equation}
The first (second) term in square brackets depends only on sum (difference) of
$x$ and $x'$. This allows us to integrate over the difference (sum) of these
variables. After some shifts of variables the remaining integral takes the form
\begin{equation}
 \Omega_\epsilon
  = -\frac{W L \epsilon^2}{2}
    \int\limits_0^1 du\, u \sin\frac{\pi u}{2} \int\limits_{-\infty}^\infty
    \frac{dy \cosh(\phi y)}{\cosh^2(\pi y) - \sin^2(\pi u/2)}
  = -\frac{W L \epsilon^2}{2 \sin(\phi/2)} \int\limits_0^1 du\, u\,
    \frac{\sin(\phi u/2)}{\cos(\pi u/2)}.
 \label{uint}
\end{equation}
The last expression is the result of $y$ integration. It can be performed, e.g.,
by closing the integration contour and summing up residues in the upper
half-plane of imaginary $y$. In order to make this sum convergent, one has to
add a weak damping factor by an infinitesimal imaginary shift of $\phi$.

We proceed with the last integral in Eq.\ (\ref{uint}) by representing
$1/\cos(\pi u/2)$ as a Fourier series
\begin{equation}
 \Omega_\epsilon
  = \frac{2 W L \epsilon^2}{\sin(\phi/2)} \frac{\partial}{\partial \phi}
    \int\limits_0^1 du\, \cos\frac{\phi u}{2}
    \sum_{n = 0}^\infty (-1)^n \cos[\pi u (n + 1/2)].
\end{equation}
Convergence of this Fourier series should also be justified by a proper damping
factor. This does not change the final result of the calculation hence we omit
such extra factors for simplicity. Performing the integration over $u$ we obtain
\begin{equation}
 \Omega_\epsilon
  = \frac{2 W L \epsilon^2}{\sin(\phi/2)} \frac{\partial}{\partial \phi}
    \cos\frac{\phi}{2} \sum_{n = 0}^\infty \left[
      \frac{1}{\pi (2n + 1) + \phi} + \frac{1}{\pi (2n + 1) - \phi}
    \right].
\end{equation}
The sum over $n$ diverges logarithmically. However, this divergence is
independent of $\phi$ and hence does not influence any observable quantities,
which are expressed as derivatives of the free energy. We can easily get rid of
the divergent part by subtracting a similar sum over $n$ with $\phi = 0$. This
yields the final result
\begin{multline}
 \Omega_\epsilon
  = \frac{2 W L \epsilon^2}{\sin(\phi/2)} \frac{\partial}{\partial \phi}
    \cos\frac{\phi}{2} \sum_{n = 0}^\infty \left[
      \frac{1}{\pi (2n + 1) + \phi} + \frac{1}{\pi (2n + 1) - \phi}
      -\frac{2}{\pi (2n + 1)}
    \right] \\
  = -\frac{W}{\pi L} \frac{(\epsilon L)^2}{\sin\frac{\phi}{2}}
    \frac{\partial}{\partial \phi} \left\{
      \cos\frac{\phi}{2} \left[
        \psi \left( \frac{\pi + \phi}{2\pi} \right)
        +\psi \left( \frac{\pi - \phi}{2\pi} \right)
        +4 \ln 2 + 2 \gamma
      \right]
    \right\}.
\end{multline}
Here $\psi$ is the digamma function and $\gamma$ is the Euler-Mascheroni
constant. The last expression yields Eq.\ (\ref{Oe_res}) of the main text (where
we drop the unobservable constant).

\section{Derivation of the sigma model}
\label{AppSigma}

In order to carry out a parametrically controlled derivation of the sigma model,
it is convenient to consider a modified problem with $n \gg 1$ flavours
of Dirac fermions. To perform the disorder average of the free energy, we also
introduce $N$ replicas. (Alternatively, one can use supersymmetry. As we will
treat the sigma model perturbatively, the two approaches are fully equivalent.)

The derivation of the sigma model starts with the fermionic action generating
the matrix Green function
(\ref{matrixG}).
\begin{equation}
 S[\phi, \phi^*]
  = \int d\mathbf{r} \sum_{a,b,\alpha} {\phi_a^\alpha}^\dagger \bigg(
      \Big\{
        i0 \Lambda_z - \bm{\sigma}\mathbf{p} - \sqrt{z} \sigma_x \big[
          \Lambda_+ \delta(x) + \Lambda_- \delta(x - L)
        \big]
      \Big\}\delta_{ab} -V_{ab}(\mathbf{r})
    \bigg) \phi_b^\alpha.
\end{equation}
Here $\Lambda_\pm = (\Lambda_x \pm i \Lambda_y)/2$ are matrices operating in RA
space. This action is the functional of two independent Grassmann two-component
(in $\sigma$ space) vector fields $\phi$ and $\phi^*$. Lower indices, $a$ and
$b$, refer to flavours while the upper index $\alpha$ enumerates replicas.
Overall, there are $4nN$ independent Grassmann variables in the Lagrangian. The
random matrix $V_{ab}$ is symmetric, that insures the time-reversal symmetry of
the model. We assume Gaussian white-noise statistics for the matrix $V$ defined
by the correlator
\begin{equation}
 \langle V_{ab}(\mathbf{r}) V_{cd}(\mathbf{r'}) \rangle
  = \frac{2\pi\alpha_0}{n} \big[
      \delta_{ac} \delta_{bd} + \delta_{ad} \delta_{bc}
    \big] \delta(\mathbf{r} - \mathbf{r}').
 \label{corr}
\end{equation}

Using the time-reversal symmetry, we rewrite the action in terms of the single
four-component field $\psi$ (and its charge-conjugate version $\bar\psi$, that
is linearly related to $\psi$):
\begin{equation}
 \psi
  = \frac{1}{\sqrt{2}} \begin{pmatrix}
      \phi \\
      i \sigma_y \phi^*
    \end{pmatrix},
 \qquad
 \bar\psi
  = i \psi^T \sigma_y \tau_x
  = \frac{1}{\sqrt{2}} \Big(
      \phi^\dagger,\; i \phi^T \sigma_y
    \Big).
\end{equation}
This introduces an additional particle-hole (PH) structure of the fields. Pauli
matrices operating in PH space are denoted by $\tau_{x,y,z}$. Bar denotes the
charge conjugation operation which has two important properties: $\bar\psi_1
\psi_2 = \bar\psi_2 \psi_1$ and $(\psi_1 \bar\psi_2)^T = \tau_x \sigma_y \psi_2
\bar\psi_1 \sigma_y \tau_x$. The action takes the following form in terms of
$\psi$:
\begin{equation}
 S[\psi]
  = \int d\mathbf{r} \sum_{a,b,\alpha} \bar\psi_a^\alpha \bigg(
      \Big\{
        i0 \Lambda_z - \bm{\sigma}\mathbf{p} - \sqrt{z} \sigma_x \big[
          \rho_+ \delta(x) + \rho_- \delta(x - L)
        \big]
      \Big\}\delta_{ab} - V_{ab}(\mathbf{r})
    \bigg) \psi_b^\alpha.
\end{equation}
In this expression we have introduced the notation $\rho_\pm = (\Lambda_x \tau_z
\pm i\Lambda_y)/2$.

Now we are ready to average $e^{-S}$ over the Gaussian disorder distribution
with the correlator (\ref{corr}). This yields an effective action with the
quartic term. Using the above-mentioned properties of charge conjugation, we
recast the action in the form
\begin{equation}
 S[\psi]
  = \int d\mathbf{r} \left[
      \sum_{a,\alpha} \bar\psi_a^\alpha \Big\{
        i0 \Lambda_z - \bm{\sigma}\mathbf{p} - \sqrt{z} \sigma_x \big[
          \rho_+ \delta(x) + \rho_- \delta(x - L)
        \big]
      \Big\} \psi_a^\alpha
      +\frac{2\pi\alpha_0}{n} \sum_{a,b,\alpha,\beta} \mathop{\mathrm{Tr}}
        \psi_a^\alpha \bar\psi_a^\beta \psi_b^\beta \bar\psi_b^\alpha
    \right].
\end{equation}
Next, we decouple the quartic term introducing an auxiliary $8N \times 8N$
matrix $R$ by the Hubbard-Stratonovich transformation. This yields the action
\begin{equation}
 S[R, \psi]
  = \int d\mathbf{r} \left[
      \frac{n \gamma^2}{8\pi\alpha_0} \mathop{\mathrm{Tr}} R^2
      + \sum_{a,\alpha, \beta} \bar\psi_a^\alpha\bigg(
        i \gamma R_{\alpha\beta} - \Big\{
          \bm{\sigma}\mathbf{p} + \sqrt{z} \sigma_x \big[
            \rho_+ \delta(x) + \rho_- \delta(x - L)
          \big]
        \Big\}\delta_{\alpha\beta}
      \bigg) \psi_a^\beta
    \right].
\end{equation}
Parameter $\gamma$ is an arbitrary number at this stage, its value will be fixed
later. Matrix $R_{\alpha\beta}$ couples to the product $\sum_a \psi_a^\alpha
\bar\psi_a^\beta$. This allows us to impose the corresponding symmetry
constraint on the matrix $R$: $R = \sigma_y \tau_x R^T \sigma_y \tau_x$.
Finally, we integrate out the fermionic fields and obtain the action operating
with the matrix $R$ only,
\begin{equation}
 S[R]
  = \frac{n}{2} \mathop{\mathbf{Tr}} \left(
      \frac{\gamma^2 R^2}{4 \pi \alpha_0} - \ln \bigg\{
        i \gamma R - \bm{\sigma}\mathbf{p} - \sqrt{z} \sigma_x \big[
            \rho_+ \delta(x) + \rho_- \delta(x - L)
        \big]
      \bigg\}
    \right).
 \label{trlog}
\end{equation}
The bold `$\mathbf{Tr}$' symbol implies the full operator trace including
integration over space coordinates.

Derivation of the sigma model proceeds with the saddle-point analysis of the
action (\ref{trlog}) in the absence of the source field $z$. We first look for a
diagonal and spatially constant matrix $R$ minimizing the action. The
saddle-point equation is identical to the self-consistent Born approximation
(SCBA) equation for the self energy $-i \gamma R$,
\begin{equation}
 - i \gamma R
  = 2 \pi \alpha_0 \int \frac{d \mathbf{p}}{(2\pi)^2} \big(
      i \gamma R - \bm{\sigma} \mathbf{p}
    \big)^{-1}.
\end{equation}
We fix $\gamma$ to be the imaginary part of the SCBA self energy, $\gamma =
\Delta e^{-1/\alpha_0}$ with $\Delta$ being ultraviolet energy cut-off (band
width). Then the saddle-point configuration for the matrix $R$ is simply $R =
\Lambda_z$. This fixes the boundary conditions for the matrix $R$ at the
contacts. Since the leads are very good metals and fluctuations of $R$ are
strongly suppressed there, $R = \Lambda_z$ for $x < 0$ and $x > L$.

The matrix $R = \Lambda_z$ is not the only saddle point of the action
(\ref{trlog}). Other configurations minimizing the action can be obtained by
rotations $R = T^{-1} \Lambda_z T$ with any matrix $T$ which commutes with
$\bm{\sigma} \mathbf{p}$ and preserves the constraint $R = \sigma_y \tau_x
R^T \sigma_y \tau_x$. Matrix $T$, and hence $R$, is trivial in $\sigma$
space. This allows us to reduce the dimension of $R$ to $4N \times 4N$ operating
in $\Lambda$, $\tau$, and replicas only. The saddle manifold generated by
matrices $T$ is $O(4N)/O(2N) \times O(2N)$.

Let us now restore the source term in the action and establish boundary
conditions for $R$. The matrix $R$ has a jump at the interfaces with the leads
due to the delta functions in the action (\ref{trlog}). However, we can
eliminate these jumps by a proper gauge transformation. Let us perform a
rotation $R = A \tilde R A^{-1}$ with an $x$-dependent matrix $A$. The action
acquires the following form in terms of $\tilde R$:
\begin{equation}
 S[\tilde R]
  = \frac{n}{2} \mathop{\mathbf{Tr}} \left[
      \frac{\gamma^2 \tilde R^2}{2 \pi \alpha_0} - \ln \left(
        i\gamma \tilde R - \bm{\sigma}\mathbf{p} + i\sigma_x A^{-1} \left\{
          \frac{\partial A}{\partial x} +i \sqrt{z} \big[
            \rho_+ \delta(x) + \rho_- \delta(x - L)
          \big] A
        \right\}
      \right)
    \right].
\end{equation}
The source field drops from this action if we choose $A$ such that the
expression in curly braces vanishes. This yields
\begin{equation}
 A
  = \begin{cases}
      1, & x < 0, \\
      1 - i \sqrt{z} \rho_+, & 0 < x < L, \\
      \big( 1 - i \sqrt{z} \rho_- \big)
      \big( 1 - i \sqrt{z} \rho_+ \big), & x > L.
    \end{cases}
\end{equation}
Note that the matrix $\tilde R$, defined with the help of the above matrix $A$,
fulfils the condition $\tilde R = \tau_x \tilde R^T \tau_x$. Since
delta functions disappear from the action, we can infer that $\tilde R$ is
continuous at the interfaces with the leads. In the left lead we have $R =
\tilde R = \Lambda_z$. This is the left boundary condition for the matrix
$\tilde R$. The right boundary condition is fixed by the identities $\tilde R =
A^{-1} R A$ and $R = \Lambda_z$ for $x > L$. This yields
\begin{equation}
 \tilde R(L)
  = (1 - 2z)\Lambda_z + i z^{3/2} \Lambda_x + \sqrt{z}(2 - z)\Lambda_y \tau_z.
\end{equation}
We can further simplify this bulky expression by performing a constant rotation
$\tilde R = B^{-1} Q B$ with the matrix
\begin{equation}
 B
  = \frac{\tau_z - \tau_y}{2\sqrt{2}} \Big[
      (1 - z)^{-1/4} (1 + \Lambda_z \tau_z)
      -i(1 - z)^{1/4} (1 - \Lambda_z \tau_z)
    \Big].
\end{equation}
After such a rotation the action and boundary conditions become
\begin{gather}
 S[Q]
  = \frac{n}{2} \mathop{\mathbf{Tr}} \left[
      \frac{\gamma^2 Q^2}{2 \pi \alpha_0} - \ln \left(
        i\gamma Q - \bm{\sigma}\mathbf{p}
      \right)
    \right], \label{trlogQ}\\
 Q(0)
  = \Lambda_z,
 \qquad
 Q(L)
  = \Lambda_z \cos\phi + \Lambda_x \sin\phi.
\end{gather}
Thus we have reduced the boundary conditions to the form (\ref{boundary}). The
matrix $B$ is chosen such that $B^T B = \tau_x$. Hence the matrix $Q$ obeys the
symmetry constraint $Q = Q^T$.

The last step of the sigma-model derivation is the gradient expansion in Eq.\
(\ref{trlogQ}). This expansion is straightforward for the real part of the
action \cite{footnote-top-term}
\begin{equation}
 \mathop{\mathrm{Re}} S[Q]
  = -\frac{n}{4} \mathop{\mathbf{Tr}}
      \ln \big(
        i\gamma Q - \bm{\sigma}\mathbf{p}
      \big) \big(
        -i\gamma Q - \bm{\sigma}\mathbf{p}
      \big)
  = -\frac{n}{4} \mathop{\mathbf{Tr}}
      \ln \big(
        p^2 + \gamma^2 + \gamma \bm{\sigma} \nabla Q
      \big)
  \simeq \frac{n}{16 \pi} \mathop{\mathrm{Tr}} (\nabla Q)^2.
\end{equation}
The Drude conductivity of the two-dimensional sample with $n$ flavours of
massless Dirac fermions at the Dirac point is $(n/\pi) (e^2/h)$. With the
dimensionless conductivity $\sigma = n/\pi$, we finally obtain Eq.\
(\ref{sigma}) supplemented by the boundary conditions (\ref{boundary}).

\twocolumngrid


\begin{thebibliography}{99}

\bibitem{Geim07}
A.\ K.\ Geim and K.\ S.\ Novoselov, Nature Materials \textbf{6}, 183 (2007).

\bibitem{RMP07}
A.\ H.\ Castro Neto, F.\ Guinea, N.\ M.\ R.\ Peres, K.\ S.\ Novoselov, and
A.\ K.\ Geim, Rev.\ Mod.\ Phys.\ \textbf{81}, 109 (2009).

\bibitem{Katsnelson}
M.\ I.\ Katsnelson, Eur.\ Phys.\ J.\ B \textbf{51}, 157 (2006).

\bibitem{Dorokhov83}
O.\ N.\ Dorokhov, Zh.\ Eksp.\ Teor.\ Fiz.\ \textbf{85}, 1040 (1983) [Sov.\
Phys.\ JETP \textbf{58}, 606 (1983)].

\bibitem{Tworzydlo06}
J.\ Tworzydlo, B.\ Trauzettel, M.\ Titov, A.\ Rycerz, and C.\ W.\ J.\ Beenakker,
Phys.\ Rev.\ Lett.\ \textbf{96}, 246802 (2006).

\bibitem{Ludwig07}
S.\ Ryu, C.\ Mudry, A.\ Furusaki, and A.\ W.\ W.\ Ludwig, Phys.\ Rev.\ B
\textbf{75}, 205344 (2007).

\bibitem{Beenakker08rev}
C.\ W.\ J.\ Beenakker, Rev.\ Mod.\ Phys. \textbf{80}, 1337 (2008).

\bibitem{BeenakkerRMP}
C.\ W.\ J.\ Beenakker, Rev.\ Mod.\ Phys.\ \textbf{69}, 731 (1997).

\bibitem{Morpurgo}
H.\ B.\ Heersche P.\ Jarillo-Herrero, J.\ B.\ Oostinga, L.\ M.\ K.\ Vandersypen,
A.\ F.\ Morpurgo, Nature \textbf{446}, 56 (2007).

\bibitem{Danneau}
R.\ Danneau, F.\ Wu, M.F.\ Craciun, S.\ Russo, M.\ Y.\ Tomi, J.\ Salmilehto,
A.\ F.\ Morpurgo, and P.\ J.\ Hakonen, Phys.\ Rev.\ Lett.\ \textbf{100}, 196802
(2008); J.\ Low Temp.\ Phys.\ \textbf{153}, 374 (2008); Solid State Comm.
\textbf{149}, 1050 (2009).

\bibitem{Miao07} F.\ Miao, S.\ Wijeratne, Y.\ Zhang, U.\ C.\ Coskun, W.\ Bao,
and C.\ N.\ Lau, Science \textbf{317}, 1530 (2007).

\bibitem{Du08}
Xu Du, I.\ Skachko, A.\ Barker, E.\ Y.\ Andrei, Nature Nanotech.
\textbf{3}, 491 (2008).

\bibitem{KimFQHE}
K.\ I.\ Bolotin, F.\ Ghahari, M.\ D.\ Shulman,H.\ L.\  Stormer, P.\ Kim, Nature
(London) \textbf{462}, 196 (2009).

\bibitem{Bardarson07}
J.\ H.\ Bardarson, J.\ Tworzyd\l o, P.\ W.\ Brouwer, and C.\ W.\ J.\ Beenakker,
Phys.\ Rev.\ Lett.\ \textbf{99}, 106801 (2007).

\bibitem{Titov07}
M.\ Titov, Europhys.\ Lett.\ \textbf{79}, 17004 (2007).

\bibitem{Schuessler09}
A.\ Schuessler, P.\ M.\ Ostrovsky, I.\ V.\ Gornyi, and A.\ D.\ Mirlin, Phys.\
Rev.\ B \textbf{79}, 075405 (2009).

\bibitem{Novoselov05}
K.\ S.\ Novoselov, A.\ K.\ Geim, S.\ V.\ Morozov, D.\ Jiang, M.\ I.\ Katsnelson,
I.\ V.\ Grigorieva, S.\ V.\ Dubonos, and A.\ A.\ Firsov, Nature (London)
\textbf{438}, 197 (2005).

\bibitem{OurPapers}
P.\ M.\ Ostrovsky, I.\ V.\ Gornyi, and A.\ D.\ Mirlin, Phys.\ Rev.\ Lett.\
\textbf{98}, 256801 (2007); Eur.\ Phys.\ J.\ Spec.\ Top.\ \textbf{148}, 63
(2007).

\bibitem{Nomura07}
K.\ Nomura, M.\ Koshino, and S.\ Ryu, Phys.\ Rev.\ Lett.\ \textbf{99}, 146806
(2007).

\bibitem{San-Jose07}
P.\ San-Jose, E.\ Prada, and D.\ S.\ Golubev, Phys.\ Rev.\ B \textbf{76}, 195445
(2007).

\bibitem{Lewenkopf08}
C.\ H.\ Lewenkopf, E.\ R.\ Mucciolo, and A.\ H.\ Castro Neto, Phys.\ Rev.\
Lett.\ \textbf{77}, 081410R (2008).

\bibitem{Tworzydlo08}
J.\ Tworzydlo, C.\ W.\ Groth, and C.\ W.\ J.\ Beenakker, Phys.\ Rev.\ B
\textbf{78}, 235438 (2008).

\bibitem{Marcus08}
L.\ DiCarlo, J.\ R.\ Williams, Yiming Zhang, D.\ T.\ McClure, C.\ M.\ Marcus,
Phys.\ Rev.\ Lett.\ \textbf{100}, 156801 (2008).

\bibitem{Nazarov95}
Yu.\ V.\ Nazarov, Phys.\ Rev.\ B\ \textbf{52}, 4720 (1995).

\bibitem{Nazarov94}
Yu.\ V.\ Nazarov, Phys.\ Rev.\ Lett.\ \textbf{73}, 134 (1994).

\bibitem{Zhang05}
Y.\ Zhang, Y.-W.\ Tan, H.\ L.\ Stormer, and P.\ Kim, Nature (London)
\textbf{438}, 201 (2005).

\bibitem{OurQHE}
P.\ M.\ Ostrovsky, I.\ V.\ Gornyi, and A.\ D.\ Mirlin, Phys.\ Rev.\ B
\textbf{77}, 195430 (2008).

\bibitem{Savchenko}
F.\ V.\ Tikhonenko, D.\ W.\ Horsell, R.\ V.\ Gorbachev, and A.\ K.\ Savchenko,
Phys.\ Rev.\ Lett.\ \textbf{100}, 056802 (2008); D.\ W.\ Horsell, A.\ K.\
Savchenko, F.\ V.\ Tikhonenko, K.\ Kechedzhi, I.\ V.\ Lerner, V.\ I.\ Fal'ko,
Solid State Comm. \textbf{149}, 1041 (2009); F.\ V.\ Tikhonenko, A.\ A.\
Kozikov, A.\ K.\ Savchenko, R.\ V.\ Gorbachev, Phys.\ Rev.\ Lett.\ \textbf{103},
226801 (2009).

\bibitem{Ando06}
T.\ Ando, J.\ Phys.\ Soc.\ Jpn \textbf{75}, 074716 (2006).

\bibitem{Nomura06}
K.\ Nomura and A.\ H.\ MacDonald, Phys.\ Rev.\ Lett. \textbf{96}, 256602 (2006).

\bibitem{Khveshchenko}
D.\ V.\ Khveshchenko, Phys.\ Rev.\ B \textbf{75}, 241406(R) (2007).

\bibitem{OurPRB}
P.\ M.\ Ostrovsky, I.\ V.\ Gornyi, and A.\ D.\ Mirlin, Phys.\ Rev.\ B
\textbf{74}, 235443 (2006).

\bibitem{Kim}
Y.-W.\ Tan, Y.\ Zhang, H.\ L.\ Stormer, and P.\ Kim, Eur.\ Phys.\ J.\ Special
Topics, \textbf{148}, 15 (2007).

\bibitem{footnote-chiral}
An alternative possibility is that the dominant disorder is formed by resonant
scatterers preserving the chiral symmetry\cite{OurPRB, Titov10}. A detailed
analysis of the evolution from ballistics to diffusion in this model will be
published elsewhere.

\bibitem{footnote-circular}
The result (\ref{Omegaa_res}) is twice smaller than the correction obtained in
Ref.\ \onlinecite{Schuessler09} within the transfer-matrix approach. The reason
for this discrepancy is that in the latter case ultraviolet divergency was
regulated by introducing a finite correlation length in the $y$ direction only.
This resulted in the angle average $\langle \delta_y^2/\bm{\delta}^2 \rangle =
1$ being twice larger than for the isotropic model adopted in the current paper,
$\langle \delta_y^2/\bm{\delta}^2 \rangle = 1/2$.

\bibitem{Aleiner06}
I.\ L.\ Aleiner and K.\ B.\ Efetov, Phys.\ Rev.\ Lett.\ \textbf{97}, 236801
(2006).

\bibitem{AltlandZirnbauer}
M.\ R.\ Zirnbauer, J.\ Math.\ Phys.\ \textbf{37}, 4986 (1996);
A.\ Altland and M.\ R.\ Zirnbauer, Phys.\ Rev.\ B \textbf{55}, 1142 (1997).

\bibitem{MirlinEvers}
F.\ Evers and A.\ D.\,Mirlin, Rev.\ Mod.\ Phys.\ \textbf{80}, 1355 (2008).

\bibitem{footnote-top-term}
The sigma-model action contains also an imaginary $Z_2$ topological
term\cite{OurPapers} ommited in Eq.\ (\ref{sigma}). This term is crucial for
ensuring topological protection from Anderson localization at small $\sigma$.
Here we are focusing, however, on the range of large $\sigma$, where the effect
of topological term is exponentially small and can be discarded.

\bibitem{Titov10}
M.\ Titov, P.\ M.\ Ostrovsky, I.\ V.\ Gornyi, A.\ Schuessler, and
A.\ D.\ Mirlin, Phys.\ Rev.\ Lett.\ \textbf{104}, 076802 (2010).

\end{thebibliography}
\end{document}